\documentclass[12pt]{iopart}
\usepackage{graphicx}
\usepackage{graphics}
\usepackage{dcolumn}
\usepackage{bm}
\usepackage{epstopdf}

\usepackage{iopams}
\begin{document}

\title[Pressure effects on the superconducting critical temperature of Li$_{2}$B$_{2}$]{Pressure effects on the electronic structure and superconducting critical temperature of Li$_{2}$B$_{2}$}

\author{E. Mart\'inez-Guerra$^1$, F. Ort\'iz-Chi$^2$, S. Curtarolo$^3$ 
and R. de Coss$^2$}

\address{$^1$ Facultad de Ciencias F\'isico-Matem\'aticas, Universidad Aut\'onoma de Nuevo Le\'on, 
Ciudad Universitaria 66450, San Nicol\'as de los Garza, N.L., M\'exico}
\address{$^2$ Departamento de F\'isica Aplicada, Centro de Investigaci\'on y de Estudios Avanzados 
del I.P.N., Cordemex 97310, M\'erida, Yucat\'an, M\'exico}
\address{$^3$ Department of Mechanical Engineering and Materials Science, Duke University, 
Hudson Hall 27708, Durham, North Carolina, USA}

\ead{edgar.martinezgrr@uanl.edu.mx}

\begin{abstract}
We present the structural, electronic and superconducting properties 
of Li$_{2}$B$_{2}$ under pressure within the framework of the density 
functional theory. The structural parameters, electronic band structure, 
phonon frequency of the $E_{{\rm 2g}}$ phonon mode and superconducting 
critical temperature $T_{c}$ were calculated for pressures up to 20 GPa. 
We predicted that the superconducting critical temperature of Li$_{2}$B$_{2}$ 
is about 11 $K$ and this decreases as pressure increases. We found that even
though the lattice dynamics of the $E_{{\rm 2g}}$ phonon mode is similar to 
MgB$_{2}$, the reduction of the $\sigma$-band density of states at Fermi level 
and the raising of the $E_{\rm 2g}$ phonon frequency with pressure were
determinant to decrease $\lambda$ and consequently $T_{c}$.
\end{abstract}


\submitto{\JPCM}

\section{Introduction}
High $T_{c}$ superconductivity in intermetallic layered structures attracted 
much attention since the discovery of superconductivity in magnesium diboride (MgB$_{2}$) in 2001 \cite{Nagamatsu2001}. However, MgB$_{2}$ keeps being the intermetallic material with the highest superconducting temperature ($39$ K) and even is well known that this property is based on 
the strong coupling of its covalent B sigma bands with B bond-stretching modes \cite{An2001,Kortus2001,Baron2007,Floris2007,Kortus2007,Pena2002}, 
it has not been possible to predict a new diboride type to enhance it. 
On the other hand, it is widely accepted that the superconductivity on 
intercalated systems as CaC$_{6}$ is dominated by the interaction of 
free-electron-like interlayer states with soft intercalant modes. Lately, 
in this context, new proposals for the quest of layered intercalated 
superconductors are being merged. The most recent theoretical study 
reports that, by decorating graphene with Li atoms, a superconductor at a much higher temperature with respect to Ca-covered graphene is obtained \cite{Profeta2012}. The Li-Be layered compounds with a low-dimensional electronic structure and strong electron-phonon (e-ph) coupling, are another family of Li-based compounds with possible superconducting properties under high pressure \cite{Feng2008,Xu2012}. More recently, there has been considerable interest in the structural stability, electronic band structure and chemical bond of Li-B compounds under high pressures \cite{Hermann2012-1,Hermann2012-2,Peng2012}.

Layered Li-based materials with a quasi-two-dimensional electronic structure and high dynamical energy scales resulting from the Li-atom motion are potential candidates for high $T_{c}$ superconductors. In that direction, first-principles calculations using data-mining approach have predicted new hypothetical intermetallic materials \cite{Kolmogorov2006,Kolmogorov2006-2} called metal-sandwich 
(MS) structures MS1(MB) and MS2(M$_{2}$B$_{2}$), which also have $sp_{2}$-layers of boron as in MgB$_{2}$ but separated by two metal layers {\it M} \cite{Kolmogorov2006}. Particularly, LiB and Li$_{2}$B$_{2}$ resulted marginally stable under ambient conditions and they become favored over the known stoichiometric compounds under pressure \cite{Kolmogorov2006,Kolmogorov2006-2}. Both compounds resulted 
remarkly interesting since the $\sigma$- and $\pi$-bands derived from its 
boron $p_{xy}$- and $p_z$-orbitals were similars to those in MgB$_{2}$ \cite{Pena2002,Calandra2007,Liu2007} and therefore a similar superconductor mechanism based on the electron-phonon coupling between the $\sigma$- and $\pi$-bands and the $E_{\rm 2g}$ phonon mode in the boron planes was expected.

Recent calculations revealed that the electron-phonon coupling ($\lambda$) 
in Li$_{2}$B$_{2}$ was weaker than in MgB$_{2}$, estimated about $0.57$ \cite{Liu2007} in comparison to those reported for magnesium diboride ranging from $0.61$ to $1.1$ \cite{An2001,Kortus2001,Pena2002,Wang2001,Bouquet2001,Yildrim2001,Choi2002}. 
Thus, the transition temperature of Li$_{2}$B$_{2}$ was reported to be around 
7 K \cite{Liu2007}. Now it is also widely accepted that in MgB$_{2}$ the electron-phonon coupling take a large contribution from $\sigma$-bands but also a small participation from $\pi$-bands\cite{Choi2002,Pena2010}. Therefore the seemingly unimportant $\pi$-electrons at $\epsilon_{F}$ could be determinant to increase the transition temperature of Li$_{2}$B$_{2}$. Thus, an appropiate doping of Li$_{2}$B$_{2}$, in which $\pi$-electrons are increased, could result in a higher $T_{c}$ than in MgB$_{2}$. This hypothesis was tested by Calandra {\it et al.} \cite{Calandra2007} substituting Li atoms with Mg or Al. Nevertheless, the tests showed that it was difficult to induce a significant amount of $\pi$-states at $\epsilon_{F}$ with small doping because the band crossing in Li$_{2}$B$_{2}$ happens to be exactly at $\epsilon_{F}$, around $2$ eV lower than in MgB$_{2} \cite{Pena2002}$. Furthermore, even from a thermodynamical stand point, doping Li2B2 might be as difficult as it is for MgB$_{2} \cite{Curtarolo2009}$.

From the McMillan equation \cite{McMillan1968} (Eq.~\ref{eq:lambda}), Li$_{2}$B$_{2}$ would be optimistic compared to MgB$_{2}$, because Li$_{2}$B$_{2}$ (0.48 states/eV) has a higher density of states of B atoms than  MgB$_{2}$(0.39 states/eV) at $\epsilon_{F}$ and this could enhance the electron-phonon coupling constant. However, this is not enough for arising T$_{c}$. One interesting case is represented by CaBeSi \cite{Massidda2009}, a material with $\sigma$- and $\pi$-bands at the Fermi level and density of states twice higher than in MgB$_{2}$, but with a very low critical temperature T$_{c}$= $0.4$ K. Thus, while the band structure can present strong similarities with both $\sigma$- and $\pi$-bands crossing the Fermi level, the phonon structure and the e-ph interaction could differ substantially.

In this paper, we have analyzed why the T$_{c}$ of Li$_{2}$B$_{2}$(7 K) has been reported lower than for MgB$_{2}$(39 K) and if it was possible to enhance the e-ph coupling of  Li$_{2}$B$_{2}$ and consequently, the superconducting temperature by applying hydrostatic pressure. Since the in-plane motion of the boron atoms changes the boron orbitals overlap, an important electron-phonon coupling could be expected mainly for the  $\sigma$- bands at $\epsilon_{F}$ and the  $E_{\rm 2g}$ phonon mode. After a careful analysis of the pressure effects on the band structure and the $E_{\rm 2g}$ phonon frequency, we found that the electron-phonon coupling and  T$_{c}$  decrease with pressure as a result of the strong reduction of the boron-$p_{xy}$ states contribution to the density of states at Fermi level and the hardening of the $E_{\rm 2g}$ phonon frequency as pressure was applied. 

\section{COMPUTATIONAL DETAILS}

\begin{figure}
\centering
\setlength{\abovecaptionskip}{0pt}
\includegraphics*[angle=-90,scale=0.7,trim=0.0mm 30.0mm 0.0mm 0.0mm]{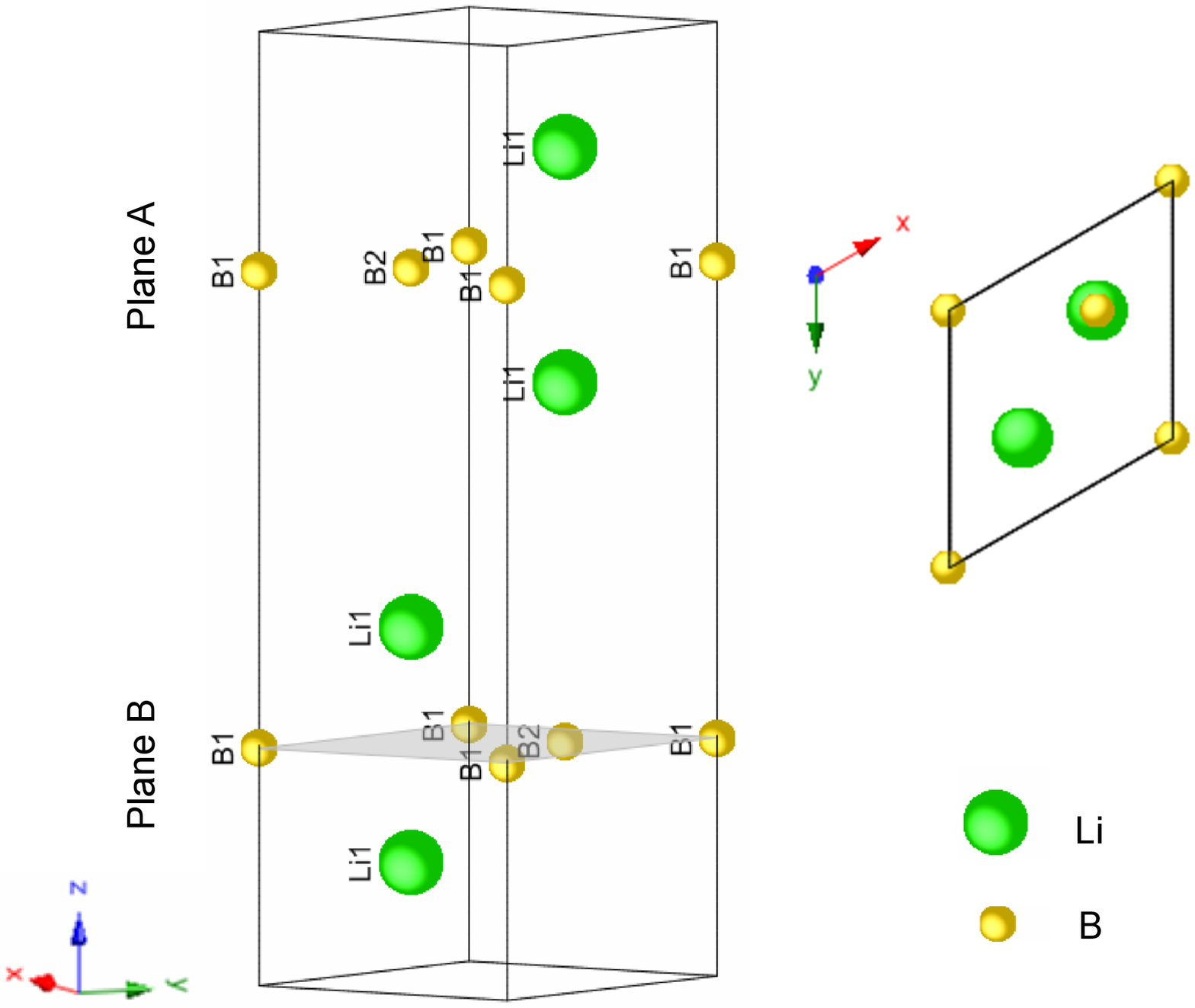}
\caption{\label{fig:Fig1} a) Crystal structure of Li$_{2}$B$_{2}$ (3D view). 
b) Hexagonal layers of boron (yellow) are separated by triangular layers of 
lithium (green) (2D view).}
\end{figure}

We studied the Li$_{2}$B$_{2}$ compound (crystal structure: MS2) as model 
of layered lithium borides. Li$_{2}$B$_{2}$ has eight atoms in the primitive 
unit cell and a space group P6$_{3}$/mmc (No.194). Wickoff positions: Li1(4f) 
($\frac{1}{3},\frac{2}{3},\frac{1}{4}-\frac{z_{m}}{4}$), B1(2b) ($0,0,\frac{1}{4}$) 
and B2(2d) ($\frac{1}{3},\frac{2}{3},\frac{3}{4}$), see Fig.~\ref{fig:Fig1}. 
The structural parameters were fully relaxed through a molecular dynamics scheme: 
$a$=$b$=$3.087$ \AA, $c$=$11.18$ {\AA}; $\alpha$=$\beta$=$90^\circ$, $\gamma$=$120^\circ$.

The total energy calculations were performed with SIESTA (Spanish Initiative for 
Electronic Simulations with Thousands of Atoms) code \cite{Soler2002,Ordejon1996} 
based on density functional theory with exchange-correlation functionals as 
parametrized by Perdew, Burke, and Ernzerhof \cite{Perdew1996} for the generalized 
gradient approximation (GGA) \cite{Langreth1983,Becke1988,Perdew1992,Perdew1993}. 
This code is implemented with pseudopotentials to describe electron-ion interactions 
and numerical atomic orbitals to expand the valence wavefunctions. Pseudopotentials 
were generated according to the Troullier-Martins procedure \cite{Troullier1991} 
using the atomic configurations 1$s^2$2$s^1$ for Li and 1$s^2$2$s^2$2$p^1$ for B 
with a core radii of 2.49 and 1.24 atomic units (a.u.) for Li and B, respectively.
As the basis set for the valence wavefunctions we have employed a double-$\xi$ basis. 
To calculate the integrals of the Khon-Sham hamiltonian, we have used a {\bf k}-space 
sampling via the Monkhorst-Pack matrix ($70\times70\times70$) with a meshcutoff of 
$280$ Ry \cite{Monkhorst1976}. All calculations were numerically converged when 
interatomic forces were less than $0.1$ meV/\AA.

Li$_{2}$B$_{2}$ phase was calculated as a function of cell volume $V$ and $c/a$ 
ratio. We performed total energy calculations for fifteen volumes and seven $c/a$ 
ratios in order to optimize both. Subsequently, the total energy was fitted to the 
Murnaghan equation of state \cite{Murnaghan1944} to obtain the equilibrium volume 
$(V_{0})$ and the $P-V$ equation of state (EOS). Thus, from the $P-V$ EOS the unit 
cell volume and $c/a$ ratio were estimated from $0$-$20$ GPa. The $E_{\rm 2g}$-phonon 
mode was calculated using the frozen-phonon model \cite{Kunk1983} that, even is a 
very simple model, it is still a suitable and useful approach \cite{Profeta2012}. 
This approach is a direct method in which the distorted Li$_{2}$B$_{2}$ is treated 
as a crystal of a lower symmetry than the non-distorted Li$_{2}$B$_{2}$. Thus, total energy calculations of distorted crystals were performed to simulate the dynamics of the 
$E_{\rm 2g}$ phonon mode. 

\begin{figure}
\centering
\setlength{\abovecaptionskip}{0pt}
\includegraphics*[angle=-90,scale=.7,trim=0mm 10.0mm 0mm 0mm, clip]{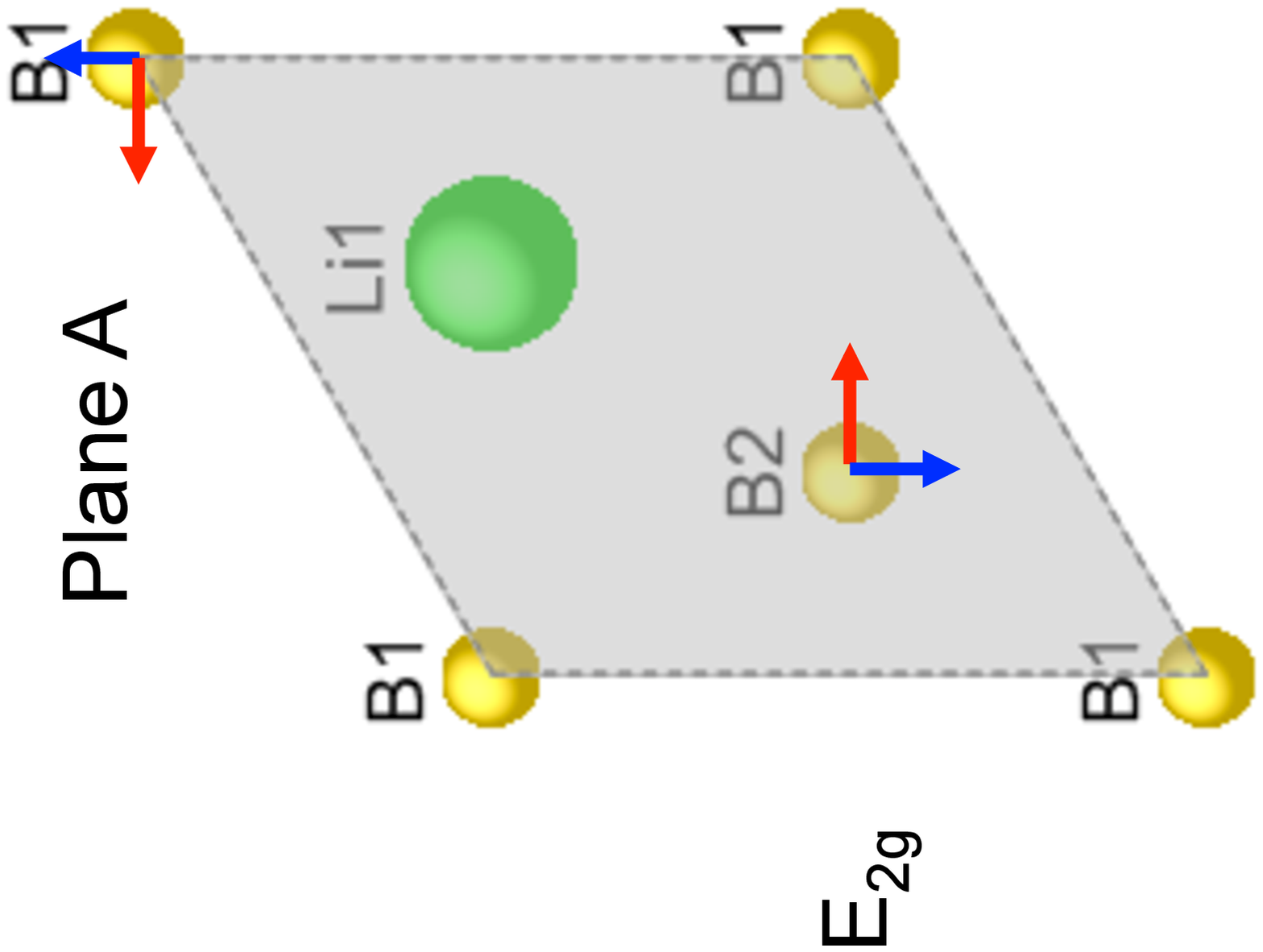}
\caption{\label{fig:Fig2} Vibration patterns of the components of the $E_{\rm 2g}$ mode.
a) $E_{\rm 2g}$(a) and b) $E_{\rm 2g}$(b) modes.}
\end{figure}

The oscillating pattern of the $E_{\rm 2g}$ 
mode corresponds to boron ions in opposite directions along $x$- or $y$-axis, 
with Li ions stationary (Fig.~\ref{fig:Fig2}). For this, we have simulated the
$E_{\rm 2g}$-phonon mode as a composition of two modes: $E_{\rm 2g}$(a) and 
$E_{\rm 2g}$(b). Figure 2 represents the vibration pattern of these modes. 
To implement the displacement pattern of these modes, we defined an adimensional 
distortion parameter $u/a$ related to the boron bond, where $u$ is the magnitude 
of the distortion and $a$ the in-plane lattice parameter, using the unit cell volume 
and $c/a$ ratio for $0$, $10$ and $20$ GPa previously optimized. These distortions 
simulated expansions and contractions of $E_{\rm 2g}$ phonon mode as a function of 
the $u$ parameter. Thus, to calculate the effective potential due to boron atoms 
oscillation we use a maximum adimensional displacement of $\left|u/a\right|$=$0.04$ 
and steps of $\Delta$=$0.01$. Eight different displacements were calculated to 
simulate the vibrational dynamics of the $E_{\rm 2g}(a)$ and $E_{\rm 2g}(b)$ 
component modes and the total energy values of these configurations were substracted 
to the total energy of non-distorted Li$_{2}$B$_{2}$ when its boron atoms are on 
its equilibrium positions. The energy differences determined the restitutive potential 
of each oscillator and they were used to solve numerically the Schr$\rm \ddot{o}$dinger 
equation and obtain the characteristic frequency of each mode.

The superconducting critical temperature of Li$_{2}$B$_{2}$ under pressure was evaluated 
using the McMillan's relation \cite{McMillan1968},
\begin{equation}
\label{eq:Tc}
T_c=\frac{\omega}{1.2}{\rm exp}\left[\frac{-1.04(1+\lambda_{\sigma})}{\lambda_{\sigma}-\mu^{*}(1+0.62\lambda_{\sigma})}\right],
\end{equation}
where $\lambda_{\sigma}$ is the coupling constant between $\sigma$-band and the $E_{\rm 2g}$ 
mode, $\omega$ is the average frequency of the $E_{\rm 2g}(a)$ and $E_{\rm 2g}(b)$ component 
modes and $\mu^*$ is an adimensional empirical value which represents the repulsive force 
between Cooper pairs. To determine the coupling constant ($\lambda_{\sigma}$) quantitatively, 
we evaluated the deformation potential $(D)$ for $E_{\rm 2g}$(a) and $E_{\rm 2g}$(b) modes 
under each distortion and used the isotropic limit from the Eliashberg 
theory \cite{Eliashberg1960} with the McMillan's formula \cite{McMillan1968},
\begin{equation}
\label{eq:lambda}
\lambda_{\sigma}=N_{B}(\epsilon_{F})\left[\frac{\hbar^2}{M\omega^2}\right]D^2,
\end{equation}
where, $\omega^2$, $D$, $N_{B}(\epsilon_{F})$ and $M$ are the average of the phonon frequency 
of the $E_{\rm 2g}$(a) and $E_{\rm 2g}$(b) modes, the deformation potential of the $\sigma$-band 
due to $E_{\rm 2g}$ mode, the density of states of boron atoms at Fermi level and total mass 
of boron atoms inside unit cell, respectively.

\section{RESULTS AND DISCUSSION}
\begin{figure}
\centering
\setlength{\abovecaptionskip}{0pt}
\includegraphics*[scale=1.50,trim=0mm 0.0mm 0mm 0mm, clip]{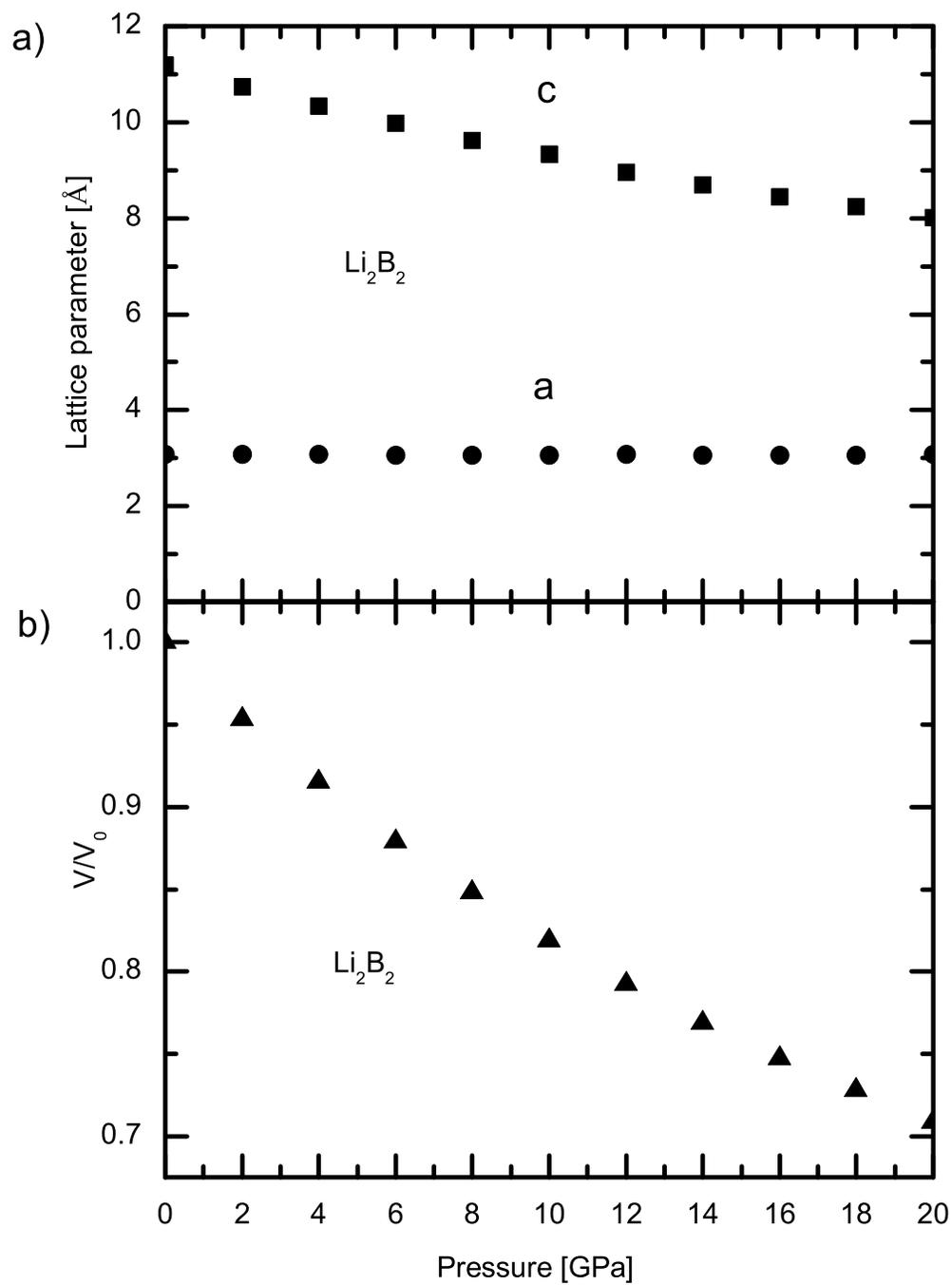}
\caption{\label{fig:Fig3} Structural parameters of Li$_{2}$B$_{2}$ as a function of pressure.
a) $a$ and $c$ lattice parameters and b) $V/V_{0}$ ratio.}
\end{figure}

In Fig.~\ref{fig:Fig3}, we show the results for the $a$ and $c$ lattice parameters, and 
the $V/V_{0}$ ratio for Li$_{2}$B$_{2}$ as a function of pressure for $0 \leq P \leq 20$ 
GPa. It can be seen that $a$ parameter decreases from 3.087 to 3.071 {\AA} while $c$ 
parameter from 11.187 to 8.0 {\AA} in the range of 20 GPa. Although both, $a$ and $c$ 
parameters decreases, it is important to note that the slope for $c$ parameter as a function 
of pressure is higher than for $a$ parameter, that is, it would be easier to compress 
Li$_{2}$B$_{2}$ through $z$-direction than in plane direction. Conversely, $V/V_{0}$ 
decreases non-linearly and Li$_{2}$B$_{2}$ could be compressed about $30\%$ when it is 
applied over 20 GPa.

\begin{figure}
\centering
\setlength{\abovecaptionskip}{0pt}
\includegraphics*[scale=0.80,trim=0mm 0.0mm 0mm 0mm, clip]{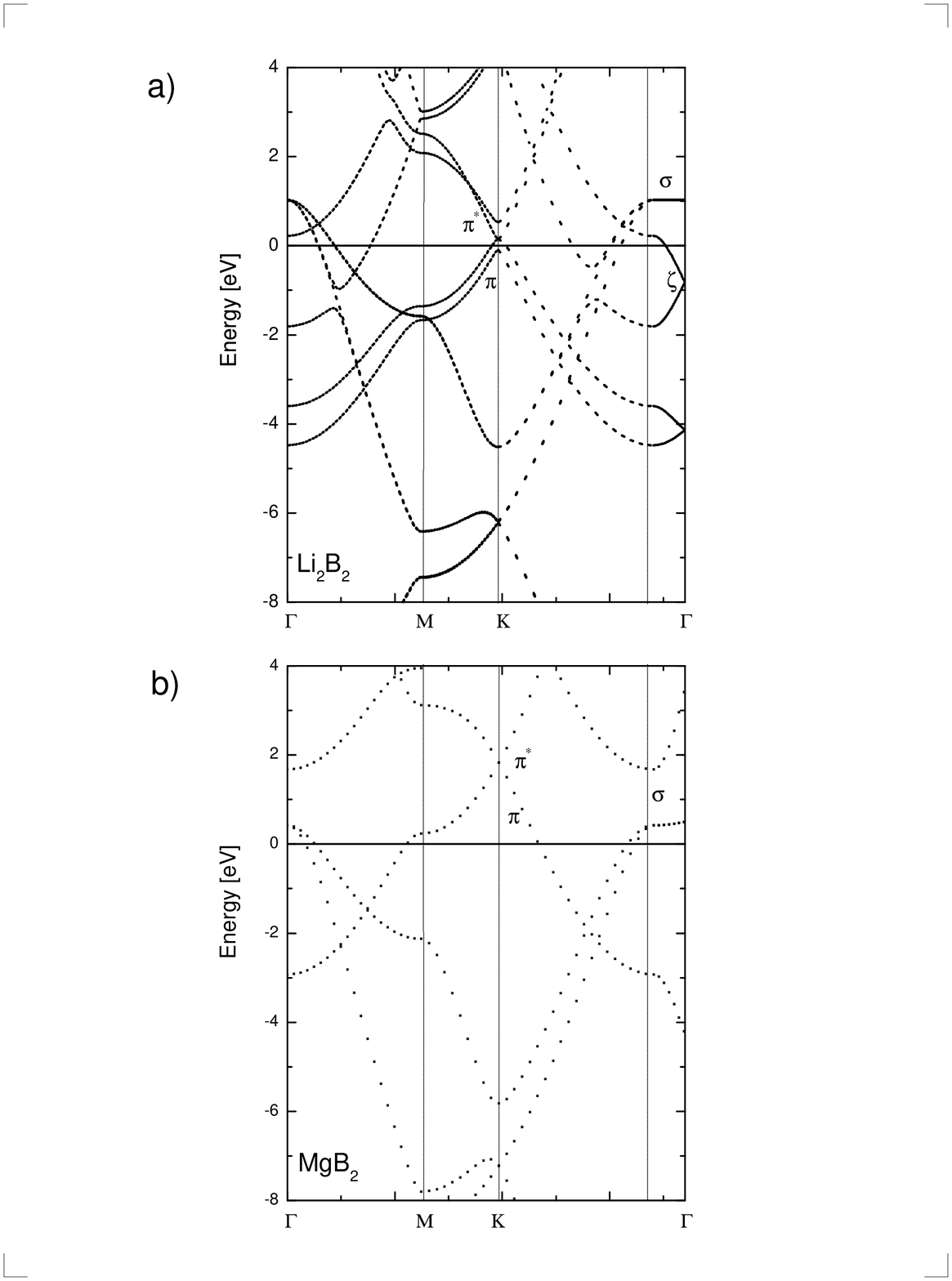}
\caption{\label{fig:Fig4} Electronic band structure at the calculated equilibrium lattice
constants for a) Li$_{2}$B$_{2}$ and b) MgB$_{2}$.}
\end{figure}

\begin{figure}
\centering
\setlength{\abovecaptionskip}{0pt}
\includegraphics*[scale=1.50,trim=0mm 0.0mm 0mm 0mm, clip]{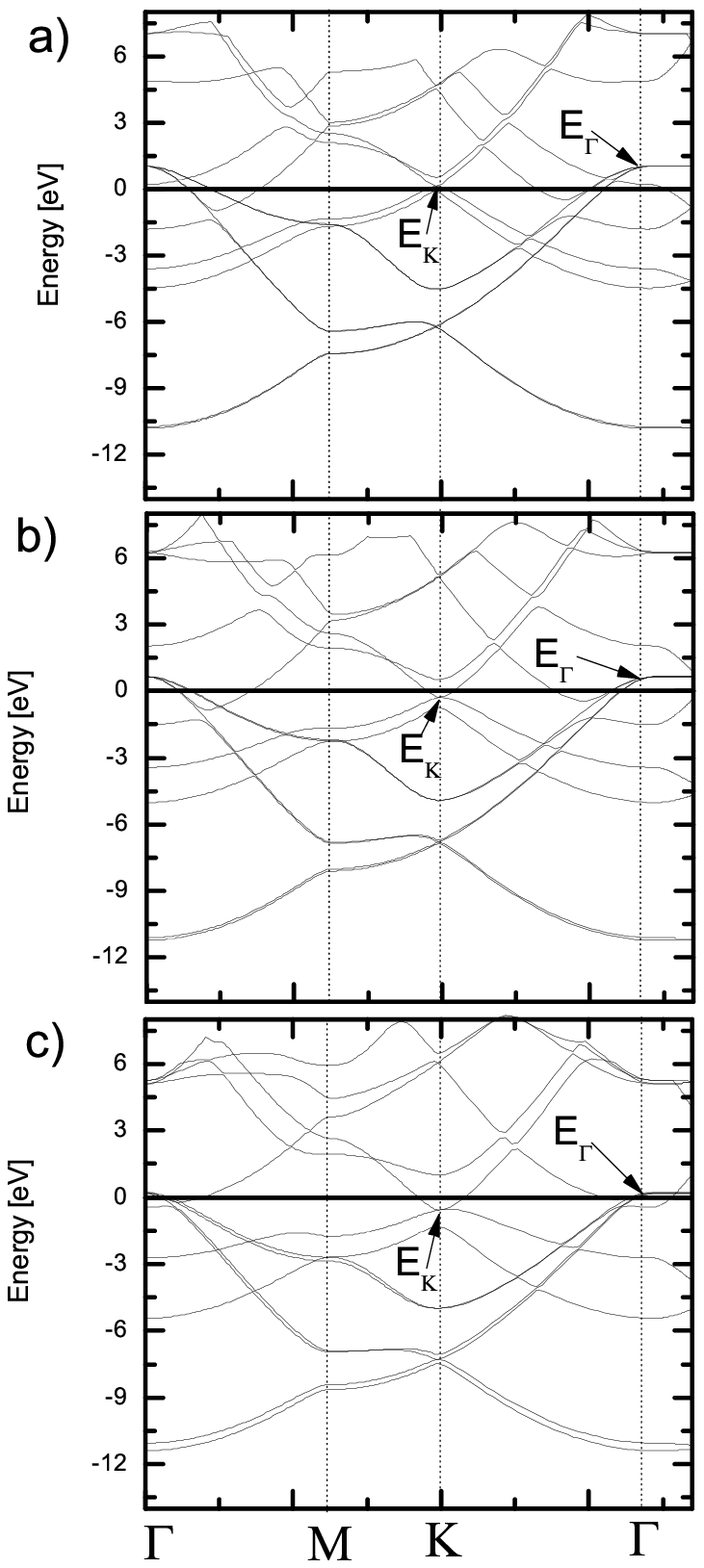}
\caption{\label{fig:Fig5} Electronic band structure of Li$_{2}$B$_{2}$ and evolution
of $\sigma$- and $\pi$-bands a) 0 GPa, b) 10 GPa and c) 20 GPa.}
\end{figure}

The band structure of MgB$_{2}$ and Li$_{2}$B$_{2}$ are shown in Fig.~\ref{fig:Fig4}. 
Both structures are as similar as it was reported before \cite{Calandra2007,Liu2007}. 
Li$_{2}$B$_{2}$ presents the $\pi$- and $\sigma$-bands which are coupled with $E_{\rm 2g}$ 
phonon mode to induce superconductivity in MgB$_{2}$. However, in Li$_{2}$B$_{2}$ the sigma band is shifted up by 0.23 eV and the pi-band is shifted down by 1.31 eV, with respect to MgB$_{2}$. This shifting produces that the $\pi$-$\pi^{*}$ crossing occurs practically at Fermi level, resulting in a 
small density of states from $\pi$-band at $\epsilon_{F}$. Hence, to evaluate the effects 
of pressure on the electronic band structure, in Fig.~\ref{fig:Fig5} the evolution of the 
$\sigma$- and $\pi$-bands with pressure is shown. From that, it can be observed the energy 
position relative to $\epsilon_{F}$ of the $\sigma$- bands at $\Gamma$ $(E_{\Gamma})$ 
and $\pi$-bands at $K$ $(E_{K})$ as a function of pressure. These results demonstrate 
that both energies, $E_{\Gamma}$ and $E_{K}$, decrease as pressure increases. 
In particular, $E_{\Gamma}$ changes from 1.03 eV at 0 GPa to 0.21 eV at 20 GPa, and 
$E_{K}$ changes from 0.11 eV at 0 GPa to $-0.56$ eV at 20 GPa. To provide an explanation 
of the energy changes of $\sigma$- and $\pi$-bands in terms of the charge redistribution 
induced by the pressure, we estimated the charge population of the valence orbitals by 
using Mulliken populations \cite{Mulliken}, as shown on Table~\ref{tab:Tab1}. A charge 
transfer from B to Li atoms is observed. The B-$p_{xy}$ and B-$p_{z}$ orbitals loose 
charge, transferring it to the Li orbitals, mainly to Li-$s$ orbital. It should be 
pointed out that the values provided in the Mulliken analysis are strongly dependent 
on the basis set, but trends observed in the charge transfer process are reliable.\\

\begin{table}
\caption{\label{tab:Tab1} Mulliken populations of valence orbitals of B and Li.}
\begin{tabular*}{\textwidth}{@{}l*{15}{@{\extracolsep{0pt plus
12pt}}l}}
\br
Atom &Pressure [GPa] &Q& $s$& $p_{x}$& $p_{y}$& $p_{z}$& $d$\\
\mr
Li &0   & 1.021  & 0.334  &  0.256  &  0.256  &  0.175  &    -  \\
   &10  & 1.070  & 0.381  &  0.279  &  0.279  &  0.131  &    -  \\
   &20  & 1.180  & 0.461  &  0.299  &  0.299  &  0.120  &    -  \\
B  &0   & 5.957  & 1.819  &  1.459  &  1.459  &  0.999  &  0.218\\
   &10  & 5.859  & 1.835  &  1.441  &  1.441  &  0.930  &  0.213\\
   &20  & 5.641  & 1.765  &  1.366  &  1.366  &  0.958  & 0.185 \\
\br
\end{tabular*}
\end{table}

\begin{figure}
\centering
\setlength{\abovecaptionskip}{0pt}
\includegraphics*[scale=1.50,trim=0mm 0.0mm 0mm 0mm, clip]{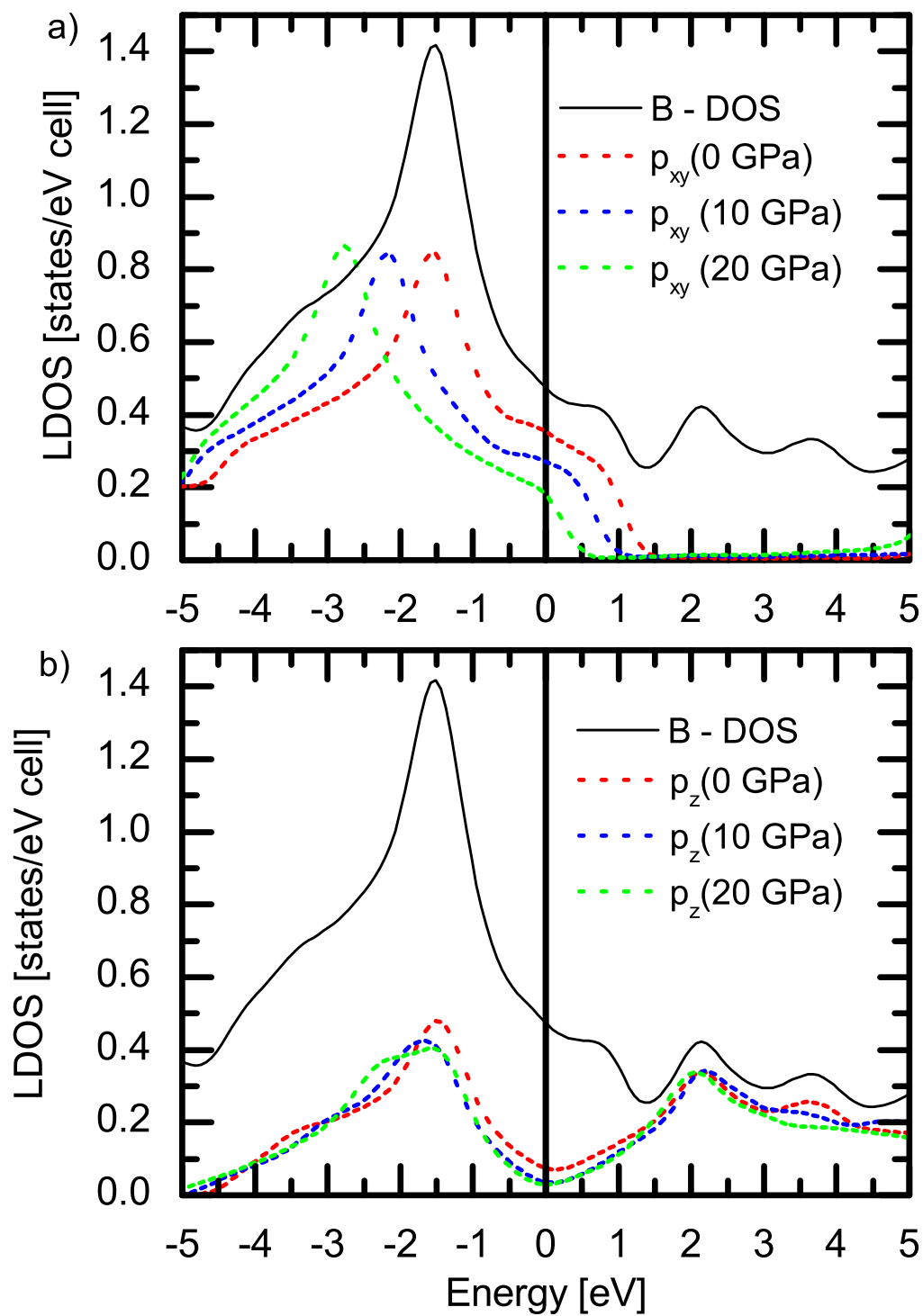}
\caption{\label{fig:Fig6} Local and partial density of states of boron atoms in 
Li$_{2}$B$_{2}$ for 0, 10 and 20 GPa. a) Partial density of states of B-$p_{xy}$ 
orbitals inside unit cell and b) partial density of states of B-$_p{z}$ orbitals 
inside unit cell.}
\end{figure}

The total and partial density of states of B, B-$p_{xy}$ and B-p$_{z}$ orbitals as a
function of pressure inside unit cell are shown in Fig.~\ref{fig:Fig6}. From that, 
it is clear that at 0 GPa the main contribution at $\epsilon_{F}$ is coming from 
the $\sigma$-band. For 0 GPa, the contributions of the B-$p_{xy}$ and B-$p_{z}$ 
orbitals to the density of states (DOS) at $\epsilon_F$ are $0.35$ and $0.075$ 
states/eV, respectively. We find that pressure induce a strong reduction of the
B-$p_{xy}$ and B-$p_{z}$ orbitals contribution to the DOS at $\epsilon_F$. Thus,
pressure of 20 GPa induces a reduction of 50\% and 37\% in the  B-$p_{xy}$ and 
B-$p_{z}$ orbitals contributions, respectively, with respect to the 0 GPa pressure 
values. These results are summarized on Table~\ref{tab:Tab2}. From these results, 
we find that the behavior of the $\sigma$- and $\pi$-bands with pressure correlates 
with the charge transfer and the reduction of states at Fermi level of the partial 
DOS corresponding to the B-$p_{xy}$ and B-$p_{z}$ orbitals. Thus, as the coupling 
constant $\lambda$ is directly proportional to $N_{B}(\epsilon_{F})$ and this one 
is weighted by contributions of B-$p_{xy}$ and B-$p_z$, the present result did not 
represent an advance in our quest to increase $\lambda$ in comparison with MgB$_2$. 
From Eq.~(\ref{eq:lambda}), we can see that the other parameters controlling $\lambda$ 
and therefore $T_{c}$ are: the phonon frequency $\omega[{E_{\rm 2g}}]$ and the 
deformation potential $D$.

\begin{table}
\caption{\label{tab:Tab2} Density of states at $\epsilon_F$ for B-$p_{xy}$ and B-$p_{z}$
orbitals at the calculated lattice constants at 0, 10 and 20 GPa.}
\begin{tabular*}{\textwidth}{@{}l*{15}{@{\extracolsep{0pt plus
12pt}}l}}
\br
Pressure [GPa]          &B($p_{xy}$) [states/eV]       &B($p_{z}$) [states/eV]\\
\mr
0  & 0.352  & 0.075\\
10 & 0.271  & 0.036\\
20 & 0.178  & 0.028\\
\br
\end{tabular*}
\end{table}

\begin{figure}
\centering
\setlength{\abovecaptionskip}{0pt}
\includegraphics*[scale=1.50,trim=0mm 0.0mm 0mm 0mm, clip]{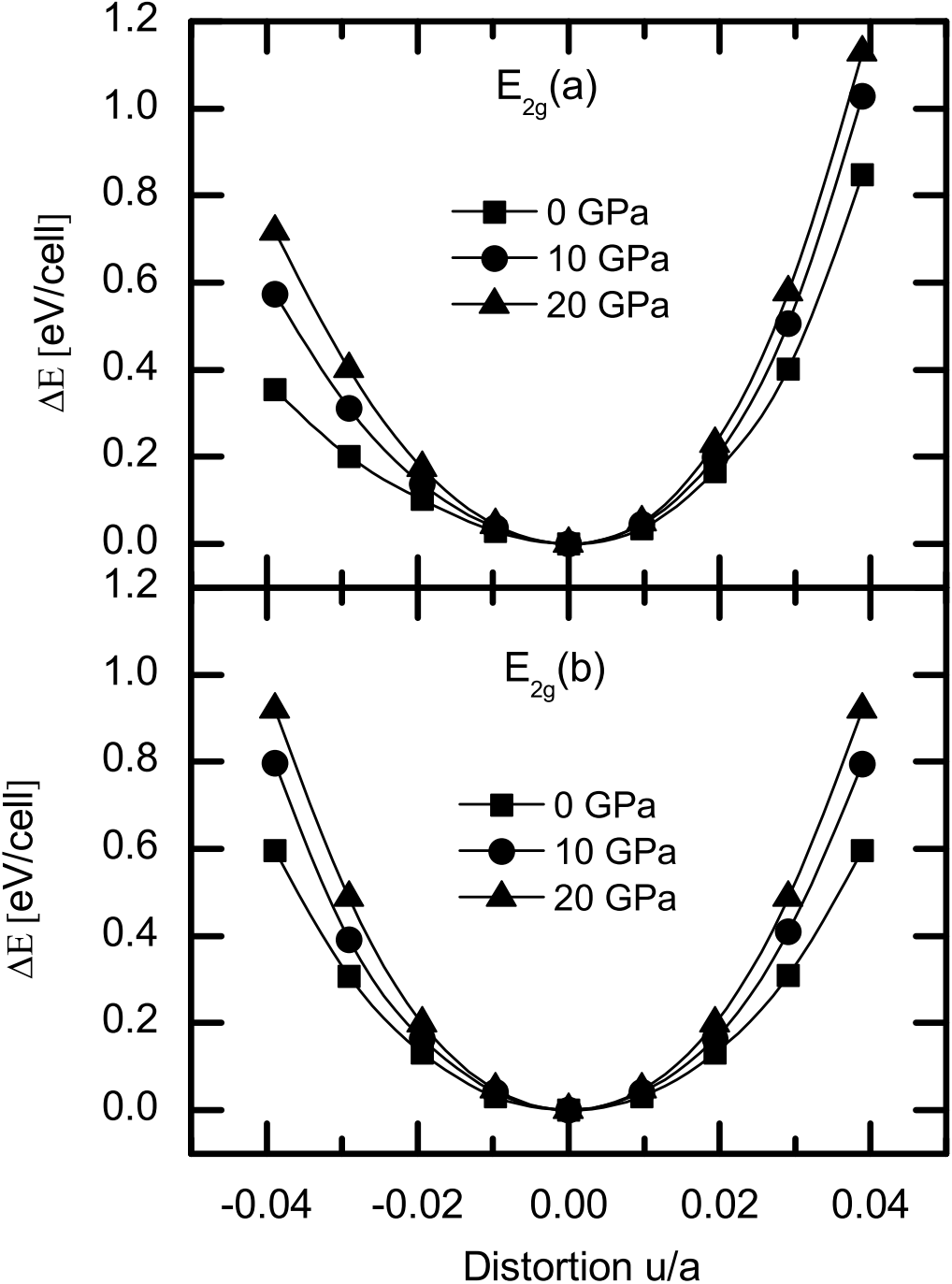}
\caption{\label{fig:Fig7} Total energy as a function of boron displacements for 
the $E_{\rm 2g}$ mode. The $E_{\rm 2g}$(a) boron in-plane mode (top) is strongly 
anharmonic while the $E_{\rm 2g}$(b) (bottom) is practically harmonic.}
\end{figure}

To further investigate $E_{\rm 2g}$ phonon frequency value, in Fig.~\ref{fig:Fig7} 
is plotted the energy as the atoms are displaced by an adimensional amount $u/a$ 
according to $E_{\rm 2g}$(a) and $E_{\rm 2g}$(b) modes. Usually, phonons are harmonic
for small displacements of the atoms with respect to equilibrium, but
anharmonic behavior is observed for large displacements. A different source of
anharmonicity take place for atomic displacement patterns with non symmetric atomic
environment, that is, for the cases when the atom experiences a different force field
for positive and negative displacements with respect to the equilibrium position.
Thus, in the present case, we find that the total-energy as a function of the
distortions for the $E_{\rm 2g}$(a) and $E_{\rm 2g}$(b) phonon modes show an
asymmetric and symmetric behavior, respectively (see Fig.~\ref{fig:Fig7}).
For the $E_{\rm 2g}$(a) mode, the distortion energy could be reasonably approximated by the 
anharmonic relation $E(u/a)=A_{2}(u/a)^{2}+A_{3}(u/a)^{3}+A_{4}(u/a)^{4}$. While for 
the $E_{\rm 2g}$(b) mode, it could be fitted to $E(u/a)=A_{2}(u/a)^{2}+A_{4}(u/a)^{4}$. 
Using the calculated well potential for each pressure and numerically solving the 
Schr$\rm \ddot{o}$dinger equation, we obtained $\omega[{E_{\rm 2g}(a)}]=79.48$ meV 
(641.06 cm$^{-1}$) when Li$_{2}$B$_{2}$ is free of pressure and 
$\omega[{E_{\rm 2g}(a)}]=93.95$ meV (757.78 cm$^{-1}$) and 
$\omega[{E_{\rm 2g}(a)}]=104.39$ meV (841.98 cm$^{-1}$) for 10 and 20 GPa, respectively. 
Therefore, the obtained results show that the hydrostatic pressure causes an increment 
in $E_{\rm 2g}$(a) phonon frequencies, and the rising does not depend linearly on the 
increasing pressure. Similarly, we obtained $\omega[{E_{\rm 2g}(b)}]=83.04$ meV 
(669.78 cm$^{-1}$) when Li$_{2}$B$_{2}$ is non-distorted and 
$\omega[{E_{\rm 2g}(b)}]=93.95$ meV (757.78 cm$^{-1}$) and 
$\omega[{E_{\rm 2g}(b)}]=104.78$ meV (845.13 cm$^{-1}$) for 10 and 20 GPa, respectively. 
Harmonic and anharmonic frequencies were estimated for $E_{\rm 2g}$(b) mode but they 
resulted to be equal and linearly dependent of pressure. Thus, because the frequency 
is directly proportional to the superconducting critical temperature, these relationship 
could be a good way to enhance $T_{c}$.

Despite that, it was quite important to quantify the deformation potential from the 
in-plane motions of the $E_{\rm 2g}$ mode changes the boron orbital overlap and 
therefore the pairing superconductor. To determine the deformation potential due to 
vibrations of the $E_{\rm 2g}$(a) and $E_{\rm 2g}$(b) modes phonons, we considered 
that only $\sigma$-band is coupled to $E_{\rm 2g}$ phonon, even that in MgB$_{2}$ 
it has been well established that $\sigma$- and $\pi$-bands are coupled with the 
$E_{\rm 2g}$ mode. We made this approximation because of two reasons: 1) lackness of 
$\pi$-states, and 2) $\sigma$-states are the main contribution to the electron-phonon
coupling in MgB$_{2}$ \cite{Kong2001,Liu2001,Golubov2002}. Thus, by estimating 
the average energy of the $\sigma$-band through $\Gamma$-$A$ path for each distortion, 
we determined the deformation potential as suggested by Khan-Allen \cite{Khan1984}, 
$D=(1/2)(d\bar{\epsilon}/du)$. As it is shown on Table~\ref{tab:Tab3}, the deformation 
potential increased slightly as a result of an increase of electron concentration when 
pressure arise. However, most remarkly is that average values of $D$ for $E_{\rm 2g}$ 
in 0$-$20 GPa range are quite close to the reported value of 12.88 eV/{\AA} for 
MgB$_{2}$ \cite{An2001} and then this contribution would be at least comparable with 
magnesium diboride.

Finally, we evaluated the electron-phonon coupling $\lambda$ of the $\sigma$-band and 
the $E_{\rm 2g}$ phonon whithin isotropic limit of the Eliashberg theory with the 
McMillan's equation and the values of $\omega[{E_{\rm 2g}}]$, $N_{B}(\epsilon_{F})$ 
and $D$ calculated previously. These results are also shown on Table~\ref{tab:Tab3}. 
From these data, it is suggested that the reduced density of states of B-$p_{xy}$ 
at $\epsilon_{F}$ and the enhanced frequency of the {$E_{\rm 2g}$} phonon mode are 
the main causes for the decreasing of $\lambda$. To establish a connection between 
$T_{c}$ and $\lambda$, we considered the strong coupling approach from Eliashberg theory, 
Eq.~\ref{eq:Tc}. By adopting the conventional value of Coulomb repulsion $\mu=0.14$ 
and two additional of $\mu^{*}=0$ and $\mu^{*}=0.1$ \cite{omar2007}, we estimated 
$T_{c}$ at different pressures. The results are summarized on Table~\ref{tab:Tab4}. 
$T_{c}$ was found to decrease from 45 to 3.6 K with a null Coulomb repulsion, from 
18.6 to 0.01 K with $\mu^{*}=0.1$ and from 10.95 to 0.07 K with $\mu^{*}=0.14$. 
With all choices of $\mu^{*}$, the resulting $T_{c}$ has a continously reduced 
trend and practically vanishes at 20 GPa.

\begin{table}
\caption{\label{tab:Tab3} Calculated values of the total density of states of boron 
atoms inside unit cell at Fermi level $N_{B}(\epsilon_{F})$, the averaged value of 
the frequencies of the components of the $E_{\rm 2g}$ phonon mode $\omega({E_{\rm 2g}})$, 
the deformation potential $D$, and the coupling constant $\lambda_{\sigma}$.}
\begin{tabular*}{\textwidth}{@{}l*{15}{@{\extracolsep{0pt plus
12pt}}l}}
\br
P[GPa] &$N_{B}(\epsilon_{F})$[$\frac{\rm states}{\rm eV}$]   &$\omega({E_{\rm 2g}})$[meV]     
       &$D_{\sigma}[\frac{\rm eV}{\rm {\AA}}]$     &$\lambda_{\sigma}$\\
\mr
0 &  0.48&  81.26&  12.88&  0.57\\
10&  0.35&  93.95&  13.17&  0.32\\
20&  0.29& 104.58&  13.21&  0.22\\
\br
\end{tabular*}
\end{table}

\begin{table}
\caption{\label{tab:Tab4} Estimated values of $T_{c}$ as a function of pressure for 
different values of Coulomb repulsion $\mu^{*}=0$, 0.1 and 0.14.}
\begin{tabular*}{\textwidth}{@{}l*{15}{@{\extracolsep{0pt plus
12pt}}l}}
\br
P[GPa]&     $\mu^{*}=0$&     $\mu^{*}=0.1$&     $\mu^{*}=0.14$\\
\mr
0 &   45&  18.6&  10.95\\
10&   12&  0.98&   0.11\\
20&  3.6&  0.01&   0.07\\
\br
\end{tabular*}
\end{table}

\section{CONCLUSIONS}
In conclusion, we have presented a first-principle investigation of the electronic 
structure, phonon frequency of the $E_{\rm 2g}$ mode, the deformation potential of 
the $\sigma$-band and the electron-phonon coupling $\lambda$ for Li$_{2}$B$_{2}$ 
under pressure from 0$-$20 GPa within the framework of density functional theory. 
We predicted a reduction of $\lambda$ from 0.57 at 0 GPa to 0.22 at 20 GPa. Even the 
lattice dynamics of the $E_{\rm 2g}$ phonon mode resulted similar to MgB$_{2}$, 
the reduced density of states of B-$p_{xy}$ at Fermi level and the raising of the 
phonon frequency corresponding to the $E_{\rm 2g}$ mode with pressure were determinant 
to decrease $\lambda$ and consequently $T_{c}$.

\section{Acknowledgments}
One of the authors (E.M.G.) gratefully acknowledges financial support from the 
Consejo Nacional de Ciencia y Tecnolog\'ia (CONACYT-M\'exico) through a posdoctoral 
scholarship (No. 55171). Also, E.M.G. would like to thank to Ramiro Quijano Qui\~nones 
and Omar de la Pe\~na Seaman for technical assistance and useful discussions. 
This research was partially funded by CONACYT-M\'exico under Grant No. 83630 and 
No. 133022.

\end{document}